\documentstyle[prc,aps,epsfig,twocolumn,eqsecnum,subfigure]{revtex}

\catcode`\@=11

\def\maketitle2{\par 
\begingroup
\let\cite\@bylinecite
\def\thefootnote{\fnsymbol{footnote}}%
\twocolumn[\@maketitle2\vskip2pc]%
\thispagestyle{plain}\@thanks
\endgroup
\def\thefootnote{\arabic{footnote}}%
\setcounter{footnote}{0}%
\let\maketitle2\relax \let\@maketitle2\relax
\let\@thanks\relax \let\@authoraddress\relax \let\@title\relax
\let\@date\relax \let\thanks\relax \let\@abstract\relax 
\let\@pacs\relax}

\def\abstract#1{\gdef\@abstract{{\par 
\bgroup
\ifdim\prevdepth=-1000pt \prevdepth0pt\fi
\hsize\columnwidth
\dimen0=-\prevdepth \advance\dimen0 by17.5pt \nointerlineskip
\small\vrule width 0pt height\dimen0 \relax}{~~}#1\egroup}}

\def\pacs#1{\gdef\@pacs{{\par 
\bgroup
\hsize\columnwidth \parindent0pt
\ifdim\prevdepth=-1000pt \prevdepth0pt\fi
\dimen0=-\prevdepth \advance\dimen0 by20pt\nointerlineskip
\egroup} PACS numbers:~#1}}

\def\@maketitle2{
\@preprint
\@title
\ifdim\prevdepth=-1000pt \prevdepth0pt\fi
\@authoraddress
\@date
\begin{list}{}{\leftmargin=0.10753\textwidth \rightmargin=\leftmargin
\itemsep=1pc\partopsep=-1pc}
\item\@abstract
\item\@pacs
\end{list}
}

\catcode`\@=12

\begin{document}
%
%
\title{Center-of-mass corrections revisited: \\
       a many-body expansion approach}
\author{Bogdan~Mihaila\thanks{electronic mail:Bogdan.Mihaila@unh.edu} and 
        Jochen~H.~Heisenberg\thanks{electronic mail:Jochen.Heisenberg@unh.edu} }
\address{Department of Physics, University of New Hampshire,
Durham, NH 03824}
\date{\today}

\abstract{
   A many-body expansion for the computation of
   the charge form factor in the center-of-mass system is proposed.
   For convergence testing purposes, 
   we apply our formalism to the case of
   the harmonic oscillator shell model, where an exact solution exists.
   We also work out the details of the calculation involving realistic 
   nuclear wave functions. Results obtained for the Argonne $v$18 two-nucleon
   and Urbana-IX three-nucleon interactions are reported.
   No corrections due to the meson-exchange charge density 
   are taken into account.
}

\pacs{21.10.Ft, 21.60.-n, 21.60.Gx, 21.60.Cs, 21.60.Jz}
\maketitle2 


%
%
\section{Introduction}

One of the successes of the shell-model picture has been the ability to
calculate self-consistent densities for nuclear ground states that 
not only reproduce experimental binding energies but also experimental
charge radii of these nuclei and generally nuclear charge densities.
The excellent agreement or remaining discrepancies have been a
cornerstone for advancing our understanding of the nuclear wave
function. In particular, the ability to predict both, heavy and light
nuclei, is taken as a confirmation of the quality of the effective
nuclear interaction used in the calculations. For that reason it is
useful to examine the accuracy with which the nuclear densities can be
calculated.

For the proper description of the scattering process one assumes
a nuclear wave function that factorizes into a nuclear center-of-mass
wave function, which is taken to be a plane wave, and an intrinsic wave
function of coordinates relative to the center-of-mass.
The difficulty lies in the ansatz of the wave function as a Slater
determinant. Such a wave function generally does not factorize into
a center-of mass wave function and a wave function for the nucleus
relative to its center-of-mass. Furthermore, for the cases where it
factorizes, the center-of-mass wave function is not a plane wave. 
While this is negligible for heavy nuclei, 
it is a significant correction for nuclei like~$^{16}$O.

This problem has been known for a long time. It can be solved exactly
for a single Slater determinant of harmonic oscillator single-particle 
wave functions.
In that case it has been shown that the wave function factorizes
with a center-of-mass wave function being a Gaussian. 
This allows us to calculate the form factor, 
i.e. the Fourier transform of the density, in the form
\begin{equation}
   F_{sd}(q) \ = \ e^{- \, \frac{1}{4} b^2 q^2 / A}  \ F_{int}(q)
\label{eq:uberal}
\end{equation}
given in terms of the harmonic oscillator length parameter $b$.
The calculation usually gives the form factor of the one-body density
labeled $F_{sd}(q)$ whereas the experiment requires the form factor
with respect to the center-of-mass, labeled $F_{int}(q)$.
Because of this exact result it has been customary to apply such a
correction also in cases where the single particle wave functions
are not harmonic oscillator wave functions and where the presence of
correlations has been substituted by an effective interaction.

An alternate way to deal with this is to calculate directly the form
factor in the center-of-mass system. 
This way the operator can be written as a series of
one-body, two-body, ..., to A-body terms. 
In this paper we first compare such an expansion with the exact result, 
for the case where such a result is available. 
We then apply the same expansion to a realistic
wave function of $^{16}$O~\cite{ref:paperone} 
and compare it to the corrections implied by equation~(\ref{eq:uberal}).
This nuclear wave function was derived for $^{16}$O
using correlations of the form $exp({\mathbf  S})$ together with the
Argonne $v$18 potential~\cite{ref:argonnev18} 
that provides an excellent fit to the nucleon-nucleon scattering 
and thus must be considered as a realistic interaction. 
Results corresponding to the inclusion of a phenomenological (Urbana-IX) 
three-nucleon interaction~\cite{ref:tnipot}
are also reported.
Thus, in this paper we hope to shed some light on the reliability of such
center-of-mass corrections.

%
%
\section{The form factor of the density}

The charge form factor at momentum transfer $\vec{q}$ is given 
in Born approximation~\cite{ref:TassieBarker} by
\begin{equation}
   F_{int}(\vec{q}) \ = \ 
   \langle 
   \phi_0 \, | \, 
          \sum_k \, f_k(q^2) \ e^{i \vec{q} \cdot \vec{r'}_k}
          \, | \, \phi_0 \rangle
   \>, 
\end{equation}
where $\phi_0$ is the translationally invariant ground state, 
$\vec{r'}_k$ the distance from the center-of-mass to the $k$th ``point" nucleon 
and $f_k(q^2)$ the nucleon form factor, which 
takes into account the finite size of the nucleon $k$. 

The center-of-mass correction has to do with the fact that the origin of the 
shell-model is not the same as the center-of-mass of the nucleus. 
Since the many-body Hamiltonian is not translationally invariant, then 
the model ground state $\Phi^{(M)}_0$ is not translationally invariant either, 
and thus can lead to incorrect description of observables, 
especially in small $A$ nuclei. 

What we need to establish is the relationship between the model quantities 
expressed in terms of the coordinates of the laboratory system 
($\vec{r}_k, \, k=1\ldots A$), 
and the intrinsic ones ($\vec{r'}_k = \vec{r}_k - \vec{R}_{cm}, \, k=1\ldots A-1$), 
measured from the center-of-mass of the nucleus
\begin{equation}
   \vec{R}_{cm} \ = \ \frac{1}{A} \, \sum_{k=1}^A \vec{r}_k
   \>.
\end{equation}
Formally, this may be viewed as a change of coordinates, 
from the coordinates of the laboratory system $\vec{r}_k$ 
to the coordinates of the center-of-mass system 
$\{ \vec{R}_{cm}, \ \vec{r'}_k \}$, 
followed by the removal of the dependence upon $\vec{R}_{cm}$ from the 
model wave function~$\Phi^{(M)}_0$, 
i.e. we have to construct the intrinsic wave function~\cite{ref:Lipkin}
\begin{equation}
   \phi^{(M)}_0 (\vec{r'}_k) \ = \ 
   \int \, G(\vec{R}_{cm}) \, \Phi^{(M)}_0(\vec{R}_{cm},\vec{r'}_k) \, d\vec{R}_{cm}
\end{equation}
independent of $\vec{R}_{cm}$, for an arbitrary function $G(\vec{R}_{cm})$. 
Note here that, in this formalism, the well-known 
Gartenhaus-Schwartz transformation~\cite{ref:GartenhausSchwartz,ref:Gibsonetal} 
corresponds to taking $G(\vec{R}_{cm}) = \delta(\vec{R}_{cm})$. 
It is clear now that the arbitrariness of the $G(\vec{R}_{cm})$ function 
causes some troubles: 
Since there is no reason to choose a particular $G(\vec{R}_{cm})$, 
it has been pointed out that the center-of-mass correction 
for a given model wave function is not uniquely defined~\cite{ref:Lipkin}.
Nevertheless, the various recipes yield 
the same result in the limit of the exact wave function of a 
free nucleus~\cite{ref:Feshbach}. 

The exact nuclear wave function $\Phi_0$ consists of two factors, 
one of which is a plane wave
in the center-of-mass coordinate, $e^{i \vec{P} \cdot \vec{R}_{cm}}$, 
the other being the intrinsic wave function $\phi_0$ of 
the relative coordinates~\cite{ref:Uberall} $\vec{r'}_k$, 
\begin{equation}
   \Phi_0(\vec{r}_1 \cdots \vec{r}_A) \ = \ 
   e^{i \vec{P} \cdot \vec{R}_{cm}} \ 
   \phi_0(\vec{r'}_1 \cdots \vec{r'}_{A-1})
   \> .
\end{equation}
For an approximate model wave function $\Phi^{(M)}_0$ however, 
all we can hope for is to be able to obtain the decomposition 
\begin{equation}
   \Phi^{(M)}_0 \ = \ 
   \phi_{cm}(\vec{R}_{cm}) \ 
   \phi^{(M)}_0(\vec{r'}_1 \cdots \vec{r'}_A)
   \> ,
\label{eq:wfcndecomp}
\end{equation}
which is approximately correct to the extent that the motion of the 
intrinsic coordinates and the center-of-mass are not correlated. 
Only then, the factorization 
\begin{equation}
   F_{sd}(\vec{q}) \ = \ 
   F_{cm}(\vec{q}) \ F_{int}(\vec{q})
\label{eq:ffdecomp}
\end{equation}
is possible. To that approximation, and  
assuming that the model provides indeed a good description of the
internal structure of the nucleus 
($\Phi_0 = \Phi^{(M)}_0$~\cite{ref:BarkerTassie}), 
equation~(\ref{eq:ffdecomp}) is valid with~\cite{ref:Feshbach}
\begin{equation}
   F_{int}(\vec{q}) \ = \ 
   \langle \Phi^{(M)}_0
           \, | \, 
           \sum_k \, f_k(q^2) \ \hat{e}_k \ 
                           e^{i \vec{q} \cdot (\vec{r}_k - \vec{R}_{cm})} 
           \, | \, 
           \Phi^{(M)}_0 \rangle
\label{eq:ffint}
\end{equation}
and
\begin{equation}
   F_{cm}(\vec{q}) \ = \ 
   \langle \Phi^{(M)}_0
           \, | \, e^{i \vec{q} \cdot \vec{R}_{cm}} \, | \, 
           \Phi^{(M)}_0 \rangle
   \> .
\end{equation}


The form factor~(\ref{eq:ffint}) can now be calculated directly 
by carrying out an expansion in terms of many-body operators
\begin{equation}
   F_{int}(\vec{q}) 
   = \sum_{k} \ f_k(q^2) \ 
              \left \langle 
              \ e^{i \vec{q} \cdot \vec{r}_k (A-1)/A}
     \prod_{m \neq k} e^{- i \vec{q} \cdot \vec{r}_m /A}
              \right \rangle 
   \>.
\label{eq:ffintstp1}
\end{equation}
Each exponential in equation~(\ref{eq:ffintstp1}) can be expressed 
in terms of the one-body operator which
we define by
\begin{equation}
   f(\vec{q} \cdot \vec{r}_m) = e^{- i \vec{q} \cdot \vec{r}_m} - 1 \>.
   \label{eq:transf}
\end{equation}
With this we write the form factor as
\begin{eqnarray}
   &&
   F_{int}(\vec{q}) 
   \ = \ \sum_{k} \ f_k(q^2) \ 
   \nonumber \\ && \times \,
                  \left \langle 
                  e^{i \vec{q} \cdot \vec{r}_k (A-1)/A}
                  \prod_{m \neq k} \,
                  \left ( 1 + f^*(\vec{q} \cdot \vec{r}_m /A) \right )
                  \right \rangle 
\end{eqnarray}
or
\begin{eqnarray}
   &&
   F_{int}(\vec{q}) 
   \ = \
        \sum_{k} \ f_k(q^2) \ 
                 \left \langle 
                 e^{i \vec{q} \cdot \vec{r}_k (A-1)/A}
                 \right \rangle 
   \nonumber \\ && 
        + \ \sum_{k} \ f_k(q^2) \
            \sum_{m \neq k} \ 
                 \left \langle 
                 e^{i \vec{q} \cdot \vec{r}_k (A-1)/A}
                 \ f^*(\vec{q} \cdot \vec{r}_m /A)
                 \right \rangle 
   \nonumber \\ && 
        + \ \frac{1}{2} \ \sum_{k} \ f_k(q^2) \ 
                 \sum_{m,n \neq k} \ 
   \nonumber \\ && \times
                    \left \langle 
                    e^{i \vec{q} \cdot \vec{r}_k (A-1)/A}
                    f^*(\vec{q} \cdot \vec{r}_m /A)
                    f^*(\vec{q} \cdot \vec{r}_n /A)
                    \right \rangle 
        + \cdots
   \>.
\label{eq:ffmbexp}
\end{eqnarray}
We intend to apply our formalism to the particular case of doubly magic nuclei ($^{16}$O). 
Thus, 
we can use the spherical symmetry of the nucleus to simplify calculations,  
in the sense that 
the form factor $F_{int}(\vec{q})$ should be spherically symmetric too, and  
we can in turn average the form factor over the directions of $\vec{q}$. 
We introduce then
\begin{equation}
   F^{(av)}_{int}(q) =  \frac{1}{4 \pi} \int F_{int}(\vec{q}) \ d \Omega_q
   \> .
\end{equation}
This allows us to write the different terms in equation~(\ref{eq:ffmbexp}) 
using the second quantization formalism, as follows
\begin{enumerate}
   \item one-body term 
       \begin{equation}
          \sum_{\alpha \beta} \ f_\alpha(q^2) \ 
                     \left \langle \alpha \, \left | \,
                     {\mathcal O}(q, \vec r_1)
                     \right | \beta \right \rangle
                     \ \mathbf{a}^{\dag}_\alpha \mathbf{a}_\beta 
          \>,
       \label{eq:1body}
       \end{equation}
       with 
       \begin{equation}
          {\mathcal O}(q, \vec r_1) 
          \ = \ 
          j_0(q r_1 (A-1)/A) \, 
          \>.
       \label{eq:1body_op}
       \end{equation}
   \item two-body term 
       \begin{eqnarray}
          &&
          \sum_{L} \ (2 L + 1) \ 
          \nonumber \\ && \times
          \sum_{\alpha \beta \gamma \delta } \ f_\alpha(q^2) \
              \left \langle \alpha \beta \, \left | \,
                    {\mathcal O}(q, \vec r_1, \vec r_2)  \, 
              \, \right | \, \gamma \delta \right \rangle  \ 
              \mathbf{a}^{\dag}_\alpha \mathbf{a}^{\dag}_\beta 
              \mathbf{a}_\delta \mathbf{a}_\gamma 
          \>,
          \nonumber \\ 
       \label{eq:2body}
       \end{eqnarray}
       with 
       \begin{eqnarray}
          {\mathcal O}(q, \vec r_1, \vec r_2) 
          & = & 
          j_L(q r_1 (A-1)/A) \, f_L(q r_2 / A) \,
          \nonumber \\ && \times \ 
          \left ( \hat{C}^{(L)}_{1} \odot \hat{C}^{(L)}_{2} \right )
          \>.
       \label{eq:2body_op}
       \end{eqnarray}
   \item three-body term 
       \begin{eqnarray}
          &&
          \sum_{L_1 L_2 L_3} i^{L_1-L_2-L_3}
               (2 L_2 + 1) (2 L_3 + 1) 
               \langle L_3 0 \, L_2 0 \, | \, L_1 0 \rangle
          \nonumber \\ && \times 
          \sum_{\alpha \beta \gamma \delta \theta \zeta} \ f_\alpha(q^2) \ 
          \left \langle \alpha \beta \gamma \, \left | \,  
               {\mathcal O}(q,\vec r_1, \vec r_2, \vec r_3) \, 
          \right | \, \delta \theta \zeta \right \rangle \ 
          \nonumber \\ && \hspace{0.5in}
          \times \ 
          \mathbf{a}^{\dag}_\alpha \mathbf{a}^{\dag}_\beta 
          \mathbf{a}^{\dag}_\gamma 
          \mathbf{a}_\zeta \mathbf{a}_\theta \mathbf{a}_\delta 
          \>,
       \label{eq:3body}
       \end{eqnarray} 
       with 
       \begin{eqnarray} 
          \lefteqn{
          {\mathcal O}(q,\vec r_1, \vec r_2, \vec r_3) }
          \nonumber \\ 
          & = &
          j_{L_1}(q r_1 (A-1)/A) \,
          f_{L_2}(q r_2 / A) \,
          f_{L_3}(q r_3 / A) \,
          \nonumber \\ && \times \,
          \left ( \hat{C}^{(L_1)}_{1} \odot 
                  \left [ \hat{C}^{(L_2)}_{2} \otimes 
                          \hat{C}^{(L_3)}_{3} 
                  \right ]^{(L_1)} 
          \right )
          \>.
       \label{eq:3body_op}
       \end{eqnarray} 
\end{enumerate}
where we have introduced $f_l(qr) = j_l(qr) - \delta_{l0}$, and 
$j_l(qr)$ and $C^{(l)}_m = \sqrt{\frac{4 \pi}{2l+1}} \, Y_{lm}(\hat{r})$ 
are the spherical Bessel functions of order $l$ and the unnormalized 
spherical harmonics of rank $l$ and component $m$, respectively. 
Greek letters label the single-particle states
$| \alpha \rangle = | n_\alpha \, (l_\alpha s_\alpha) j_\alpha m_{j_\alpha}; 
\tau_\alpha m_{\tau_\alpha} \rangle$, 
with $s = \frac {1} {2}$, $\tau = \frac {1} {2}$, $j = l \pm \frac {1} {2}$
and $m_\tau = + \frac {1} {2} ( - \frac {1} {2})$ -- for a proton 
(neutron).
As a final remark, note that the conversion to second quantization allows for 
all restrictions in the sums~(\ref{eq:ffmbexp}) to be dropped.

%
%
\section{Harmonic Oscillator Shell-Model Calculation}

Equation~(\ref{eq:ffdecomp}) is always exact if $\Phi^{(M)}_0$ is expressed in 
terms of harmonic oscillator wave functions, provided that 
the center-of-mass wave function $\phi_{cm}$ is in one given 
harmonic oscillator state. 
Then, 
the extraction of the center-of-mass coordinate can be done analytically.
Elliott and Skyrme~\cite{ref:ElliottSkyrme} have shown long time ago, 
that if the shell-model states are non-spurious,
then the center-of-mass moves in its ground state and is described by 
the 1\emph{s} harmonic oscillator wave function
\begin{equation}
   \phi_{cm}(\vec{R}_{cm}) \ = \ 
   \left ( \frac{A^3}{\pi^3 \, b^6} \right )^{\frac{1}{4}} \, 
   \exp \left [ - \frac{A \, R^2_{cm}}{2 \, b^2} \right ]
   \> ,
\label{eq:CMHO}
\end{equation}
where $b$ is the harmonic oscillator length parameter.
The center-of-mass form factor can also be evaluated explicitly 
\begin{equation}
   F_{cm}(\vec{q}) \ = \ e^{- \, \frac{1}{4} b^2 q^2 / A}
   \> .
\label{eq:ffCMHO}
\end{equation}
The correct translation-invariant form factor is thus given in terms 
of the shell-model form factor by 
\begin{equation}
   F_{int}(\vec{q}) \ = \ 
   e^{\frac{1}{4} b^2 q^2 / A} \ F_{sd}(\vec{q}) 
   \> ,
\label{eq:ffCMcorr}
\end{equation}
i.e. $F_{sd}$ must be corrected by dividing through $F_{cm}(q)$.
Note that, since the uniqueness of the procedure of carrying out 
the center-of-mass corrections has been questioned, 
the use of the equation~(\ref{eq:ffCMcorr}) 
has been suggested even in the case of a more 
general nuclear structure model~\cite{ref:Uberall}. 


We exploit the analytical nature of these results by testing 
how fast does the many-body expansion~(\ref{eq:ffmbexp}) converge. 
The shell-model wave function $\Phi^{(M)}_0$ 
for the harmonic oscillator potential is an independent particle wave function, 
represented by a simple Slater determinant of single-particle orbits.
This state is what we shall call the uncorrelated ground state $|0\rangle$.
By taking the expectation value 
in the model ground state $\Phi^{(M)}_0 \, = \, |0\rangle$, 
of the one-, two- and three-body operators
in equations~(\ref{eq:1body}), (\ref{eq:2body}) and (\ref{eq:3body}), 
the following relevant expectation values are obtained
\begin{eqnarray}
   \langle 0 \, | \,
             \mathbf{a}^{\dag}_\alpha \mathbf{a}_\beta
             \, | \, 0 \rangle
   & = & \delta_{\alpha \beta}
   \\
   \langle 0 \, | \,
             \mathbf{a}^{\dag}_\alpha \mathbf{a}^{\dag}_\beta
             \mathbf{a}_\delta \mathbf{a}_\gamma
             \, | \, 0 \rangle
   & = & \delta_{\alpha \gamma} \delta_{\beta \delta} \, - \,
         \delta_{\alpha \delta} \delta_{\beta \gamma}
   \\
   \langle 0 \, | \,
      \mathbf{a}^{\dag}_a \mathbf{a}^{\dag}_b
      \mathbf{a}^{\dag}_c
      \mathbf{a}_\zeta \mathbf{a}_\theta \mathbf{a}_\delta
      \, | \, 0 \rangle
   & = &
     \delta_{\alpha \delta} \left (
             \delta_{\beta \theta} \delta_{\gamma \zeta} \, - \,
             \delta_{\beta \zeta} \delta_{\gamma \theta} \right )
   \nonumber \\ &&
   - \,
     \delta_{\alpha \theta} \left (
             \delta_{\beta \delta} \delta_{\gamma \zeta} \, - \,
             \delta_{\beta \zeta} \delta_{\gamma \delta} \right )
   \nonumber \\ &&
   + \,
     \delta_{\alpha \zeta} \left (
             \delta_{\beta \delta} \delta_{\gamma \theta} \, - \,
             \delta_{\beta \theta} \delta_{\gamma \delta} \right )
   \> .
\end{eqnarray}
Using these results and following a straight forward but laborious calculation, 
the translation-invariant form factor for the 
harmonic oscillator shell-model can be computed completely 
up to the third-order in the many-body expansion~(\ref{eq:ffmbexp}). 
The various components involved are presented here, 
by their corresponding term of origin in the many-body expansion.
Summations over all ($nlj$) indices are implicit.
Notations are discussed in an Appendix.

\paragraph{One-body term.} 
There is only one contribution to the one-body term of $F^{(av)}_{int}(q)$
\begin{eqnarray}
   {\mathrm HO1} 
   & = & 
   f_{n l j}(q^2) \, 
   I^{(1) \, 0}_{\scriptstyle n l, n l}
   \>,
\end{eqnarray}
where ${\mathcal HO}_{n l}(r)$ are the usual radial harmonic oscillator 
wave functions. 
Note that, in the previous equation, ${\mathrm HO1}$ is actually 
the Fourier transform of the one-body density folded with the appropriate 
nucleon form factor, i.e.
\begin{eqnarray}
   {\mathrm HO1} & = & 
   f_{p}(q^2)  \ \int_0^\infty \, \rho^{\, (p)}_0(r)  \  
   j_0({\scriptstyle \frac{A-1}{A}} q r) \  r^2 \, dr 
   \nonumber \\ && 
   \ + \ 
   f_{n}(q^2) \ \int_0^\infty \, \rho^{\, (n)}_0(r) \ 
   j_0({\scriptstyle \frac{A-1}{A}} q r) \  r^2 \, dr
   \>,
   \nonumber \\
\label{eq:folden}
\end{eqnarray}
where $\rho^{\, (p)}_0(r)$ and $\rho^{\, (n)}_0(r)$ 
are the proton and neutron one-body densities, respectively, 
corresponding to the uncorrelated ground state~$| 0 \rangle$.

\paragraph{Two-body term.} 
Two components contribute to the two-body term of $F^{(av)}_{int}(q)$  
\begin{enumerate}
   \item one component corresponding to the direct contraction 
         $\delta_{\alpha \gamma} \delta_{\beta \delta}$
         \begin{eqnarray}
            {\mathrm HO2_{dr}} & = &
            f_{n_1 l_1 j_1}(q^2) 
            I^{(1) \, 0}_{\scriptstyle n_1 l_1 j_1, n_1 l_1 j_1}
            I^{(2) \, 0}_{\scriptstyle n_2 l_2 j_2, n_2 l_2 j_2}
            \>;
            \nonumber \\ 
         \end{eqnarray}

   \item one component associated with the exchange contraction 
         $\delta_{\alpha \gamma} \delta_{\beta \delta}$
         \begin{eqnarray}
            {\mathrm HO2_{ex}} 
            & = &
            f_{n_1 l_1 j_1, \, n_2 l_2 j_2}(q^2) \
            \sum_L \,  
            (2L+1) 
            \nonumber \\ && \times
            \bar I^{(1) \, L}_{\scriptstyle n_1 l_1 j_1, n_2 l_2 j_2} \, 
            \bar I^{(2) \, L}_{\scriptstyle n_1 l_1 j_1, n_2 l_2 j_2}
            \>;
         \end{eqnarray}
         where the pair of indices of the nucleon form factor $f(q^2)$ indicate
         that the two orbits denoted as $(n_1 l_1 j_1)$ and $(n_2 l_2 j_2)$ 
         have the same isospin.
\end{enumerate}

\paragraph{Three-body term.} 
The three-body term contains six contributions to $F^{(av)}_{int}$, 
out of which two are identical due to the fact that, 
in equation~(\ref{eq:3body}), 
the radial and angular parts of the operator dependent upon the coordinates
of the $2$nd nucleon are the same as the radial and angular parts of 
the operator dependent upon the coordinates of the $3$rd nucleon. 
The different components of the three-body term~(\ref{eq:3body}) 
are listed below
\begin{enumerate}
   \item \emph{term 3.1}
      ($\delta_{\alpha \delta} \delta_{\beta \theta} \delta_{\gamma \zeta}$)
         \begin{eqnarray}
            {\mathrm HO3_{1}} 
            & = &
            f_{n_1 l_1 j_1}(q^2) \ 
                I^{(1) \, 0}_{\scriptstyle n_1 l_1 j_1, n_1 l_1 j_1}
            \nonumber \\ && \times \ 
                I^{(2) \, 0}_{\scriptstyle n_2 l_2 j_2, n_2 l_2 j_2}
                I^{(2) \, 0}_{\scriptstyle n_3 l_3 j_3, n_3 l_3 j_3}
            \>;
         \end{eqnarray}
   \item \emph{term 3.2}
      ($\delta_{\alpha \delta} \delta_{\beta \zeta} \delta_{\gamma \theta}$)
         \begin{eqnarray}
            \lefteqn{
            {\mathrm HO3_{2}} \ = \ 
                - f_{n_1 l_1 j_1}(q^2)
                I^{(1) \, 0}_{\scriptstyle n_1 l_1 j_1, n_1 l_1 j_1}
            }
            \nonumber \\ && \times
                \sum_L (-1)^L (2L+1)
                ( \bar I^{(2) \, L}_{\scriptstyle n_2 l_2 j_2, n_3 l_3 j_3} )^2
            \>;
         \end{eqnarray}
   \item \emph{term 3.3}
      ($\delta_{\alpha \theta} \delta_{\beta \delta} \delta_{\gamma \zeta}$)
      is equal to \emph{term 3.6}
      ($\delta_{\alpha \zeta} \delta_{\beta \theta} \delta_{\gamma \delta}$)
         \begin{eqnarray}
            \lefteqn{
            {\mathrm HO3_{3}} \ = \ 
            {\mathrm HO3_{6}} 
            }
            \nonumber \\ 
            & = &
                - f_{n_1 l_1 j_1, \, n_2 l_2 j_2}(q^2)
                I^{(2) \, 0}_{\scriptstyle n_3 l_3 j_3, n_3 l_3 j_3}
            \nonumber \\ && \times
                \sum_L \ (2L+1)
                \bar I^{(1) \, L}_{\scriptstyle n_1 l_1 j_1, n_2 l_2 j_2}
                \bar I^{(2) \, L}_{\scriptstyle n_1 l_1 j_1, n_2 l_2 j_2}
            \>;
         \end{eqnarray}
   \item \emph{term 3.4}
      ($\delta_{\alpha \theta} \delta_{\beta \zeta} \delta_{\gamma \delta}$)
      is equal to \emph{term 3.5}
      ($\delta_{\alpha \zeta} \delta_{\beta \delta} \delta_{\gamma \theta}$)
         \begin{eqnarray}
            &&
            {\mathrm HO3_{4}} \ = \
            {\mathrm HO3_{5}} \ = \ 
                - f_{n_1 l_1 j_1, \, n_3 l_3 j_3}(q^2) \  
            \nonumber \\ && \times
                \sum_{L_2} \
                     (2 L_2 + 1) 
                \sum_{L_3} \
                     (2 L_3 + 1) 
                \sum_L \ 
                     i^{L+L_2+L_3} 
            \nonumber \\ && \times
                     \sqrt{2L+1} \langle L_2 0 \, L_3 0 \, | \, L 0 \rangle 
                     \left\{ \begin{array}{ccc}
                             L_3 & L   & L_2 \\
                             j_1 & j_2 & j_3
                     \end{array} \right\} 
            \nonumber \\ && \times
                   \bar I^{(1) \, L}_{\scriptstyle n_1 l_1 j_1, n_3 l_3 j_3}
                   \bar I^{(2) \, L_3}_{\scriptstyle n_2 l_2 j_2, n_1 l_1 j_1}
                   \bar I^{(2) \, L_3}_{\scriptstyle n_3 l_3 j_3, n_2 l_2 j_2}
            \>;
         \end{eqnarray}
\end{enumerate}

\begin{figure}
   \centering
   \epsfxsize = 3.0in
   \subfigure[$^{4}$He]
   {\epsfig{file=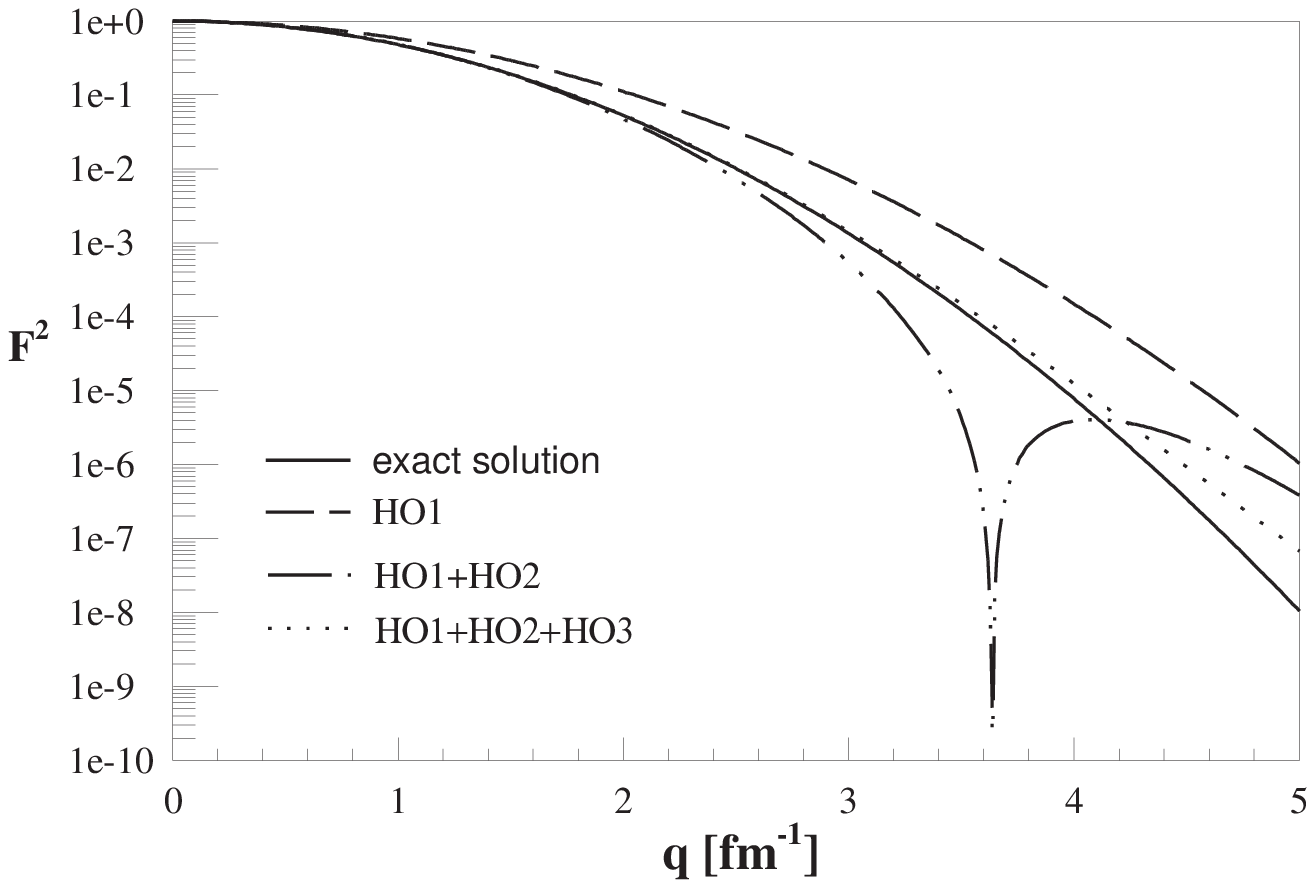,width=.50\textwidth}}
   \subfigure[$^{16}$O]
   {\epsfig{file=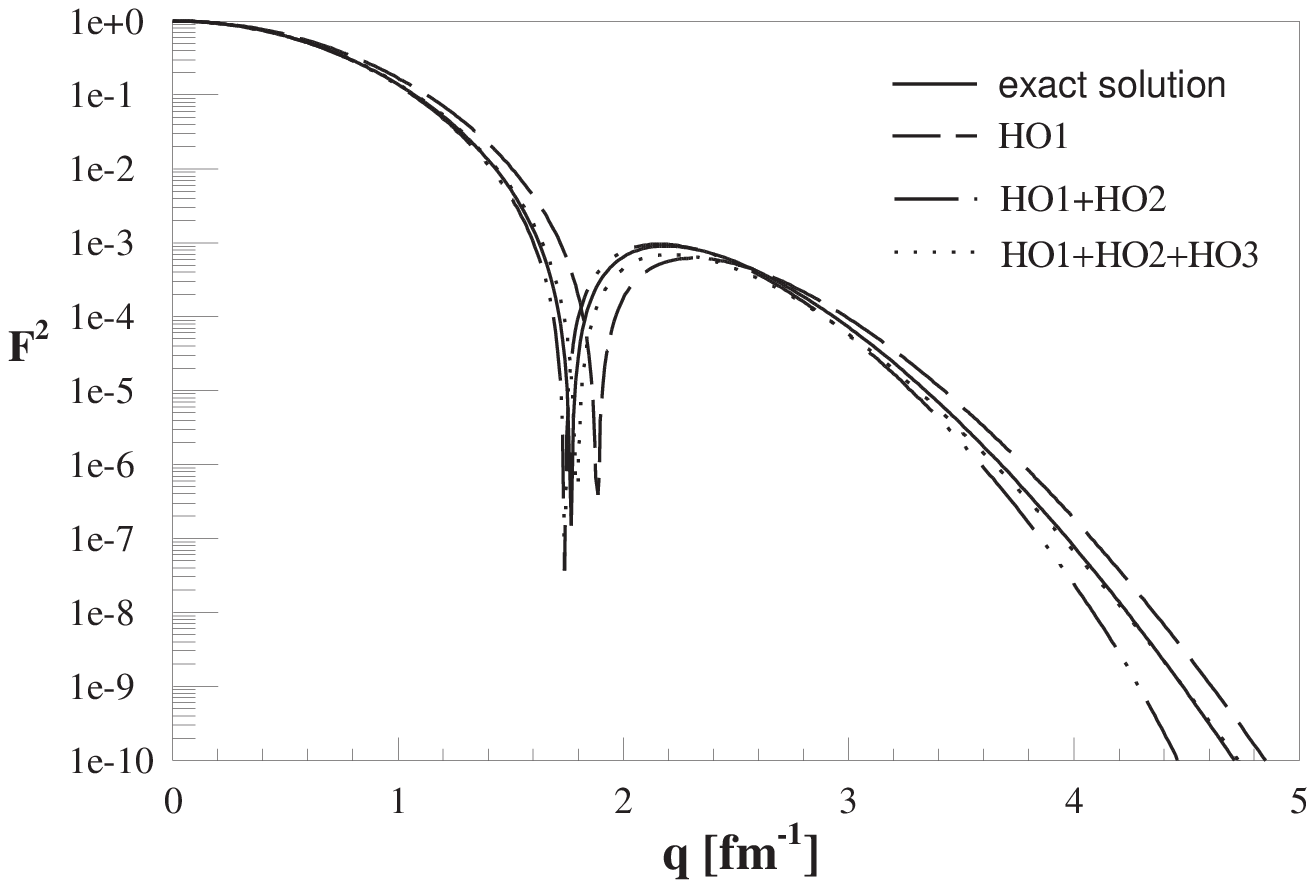,width=.50\textwidth}} 
   \caption{Convergence of the many-body expansion~(\ref{eq:ffmbexp}) 
            of the charge form factor,
            for the harmonic oscillator shell model case.}
\label{fig:fig_test_HO}
\end{figure}

In Fig.~\ref{fig:fig_test_HO} we illustrate 
the convergence of the many-body expansion, 
for the case of the $^{4}$He and $^{16}$O nuclei, respectively. 
The solid line represents the \emph{exact} form factor in the 
center-of-mass system, as given by Eq.~(\ref{eq:ffCMcorr}). 
The agreement is excellent for a momentum transfer $q \, < \, 3$ fm$^{-1}$, 
and remains reasonable good for $q$ up to 4 fm$^{-1}$. 
It is expected that the size of the contributions due to correlations
(as presented in the next section), is more important than the error made
by ignoring higher order terms in the many-body expansion (\ref{eq:ffmbexp}). 
Also, it is worthwhile mentioning that a correction expected to become increasingly 
important for high values of the momentum transfer, 
is the contribution due to the meson-exchange charge density~\cite{ref:SchiavillaETal}.
However, the inclusion of this correction is beyond 
the purpose of the present discussion. 

We conclude that truncating the calculation at the third-order gives us a good 
approximation of the center-of-mass correction for the independent-particle 
model wave function case. 
Note that leaving out the three-body term in the case of the $^{4}$He nucleus, 
would result in an unacceptable description of the form factor distribution
-- \emph{false} minima are located at a momentum transfer $q$ as low as 3.6 fm$^{-1}$ --,
whereas in the case of the $^{16}$O nucleus, the charge form factor changes very little 
by including the three-body term.
This is an indication that expression~(\ref{eq:ffCMcorr}) 
can be viewed effectively, as a $1/A$ power expansion of the charge form factor. 
Therefore, as we consider the applicability of the 
expansion~(\ref{eq:ffCMcorr}) for higher values of $A$, it appears that
we can safely drop higher-order terms in the many-body expansion and
still hope for a good description charge form factor.

To conclude our study of the convergence of the many-body 
expansion, let us investigate the influence a
given order of approximation has on the inferred 
mean square charge (rms) radius.
It is well known that in the low $q$ limit, the form factor may be be expanded 
in power series as
\begin{equation}
   F_{int}(q) \ = \ 1 \, - \, \frac{1}{6} \, q^2 \, \langle r^2 \rangle 
                      \, + \, \cdots
   \>,
\end{equation}   
and thus is a measure of the rms radius. 
Table~(\ref{tab:rms_ho}) shows the convergence of the rms radius
for the case of the $^{4}$He and $^{16}$O nuclei. 
These results show that the rms radius is little affected 
by any corrections beyond the two-body term of the expansion~(\ref{eq:ffmbexp}).
By including the three-body term in Eq.~(\ref{eq:ffmbexp}), the rms radius 
remains virtually the same in the $^{4}$He case, and changes by less 
than 1 \% in the $^{16}$O case.

%
%
\section{Realistic nuclear wave function using the $\exp(S)$ method}

We shall apply now  our formalism to the case of a more complicated 
model wave function $\Phi^{(M)}_0$ and the particular case of the 
$^{16}$O nucleus. 
As advertised, the nuclear wave function 
$\Phi^{(M)}_0 \, = \, | \tilde{0} \rangle$, 
has been obtained using the coupled cluster method 
(or the $\exp({\mathbf S})$ method) 
together with a \emph{realistic} interaction~\cite{ref:paperone}. 
The \emph{exact} correlated ground state ket wave function~$|\tilde{0}\rangle$, 
is written in terms of the uncorrelated ground state~$|0\rangle$, as
\begin{equation}
   | \tilde{0} \rangle = {\displaystyle e^{\mathbf S^{\dag}}} | 0 \rangle 
   \>.
\label{eq:corrgs}
\end{equation} 
Here, ${\rm S}^{\dag}$ is the cluster correlation operator, which may 
be decomposed in terms of  \emph{ph}-creation operators 
(${\mathbf O}^{\dag}_0$~=~${\mathbf 1}$,  
 ${\mathbf O}^{\dag}_1$~=~${\mathbf a}^{\dag}_{p_1} {\mathbf a}_{h_1}$, 
 ${\mathbf O}^{\dag}_2$~=~${\mathbf a}^{\dag}_{p_1} {\mathbf a}^{\dag}_{p_2} 
                           {\mathbf a}_{h_2} {\mathbf a}_{h_1}$),
as
\begin{equation}
   {\mathbf S}^{\dag} = \sum_{n=0}^\infty \frac{1}{n!} S_n {\mathbf O}^{\dag}_n 
   \>.
\label{eq:Sdef}
\end{equation}

The expectation value of an arbitrary operator $A$  
in the energy eigenstate~(\ref{eq:corrgs}) may be written as
\begin{equation}
   \bar{A} = 
   \langle 0 \, | \, 
             e^{\mathbf S} \, A \, e^{- \mathbf S} \, \tilde{\mathbf S}^\dag 
             \, | \, 0 \rangle 
   \>,
\end{equation}
where similarly to ${\mathbf S}^\dag$, $\tilde{\mathbf S}^\dag$ is defined by
its decomposition in terms of {\em ph}-creation operators
\begin{equation}
   \tilde{\mathbf S}^{\dag} = 
   \sum_n \frac{1}{n!} \tilde{S}_n {\mathbf O}^{\dag}_n 
   \>.
\label{eq:STdef}
\end{equation}

Therefore, the correct translation-invariant form factor is given 
by the expectation value of the operator $F_{int}$
in the correlated ground state $| \tilde{0} \rangle$. 
As we have previously~\cite{ref:paperone} worked out 
the one- and two-body densities for the ground state, 
we can apply these results to evaluate the first two terms in this expansion. 

Using the definition of the one-body density
\begin{equation}
   \rho(\vec r) 
   \ = \ 
   \sum_m 
   \langle 
       \tilde 0 \, | \, \delta(\vec r - \vec r_m) \, | \, \tilde 0 
   \rangle 
   \>,
\end{equation}
together with the identity
\begin{eqnarray}
   e^{i \vec q \cdot \vec r_k}
   \ = \
       \int \ d\vec{r} \ 
              e^{i \vec q \cdot \vec r} \       
              \delta(\vec r - \vec r_k)
   \>,
\end{eqnarray}
we can write the first term of Eq.~(\ref{eq:ffmbexp}) as
\begin{eqnarray}
   A_1 
   & = &
   f_{p}(q^2) \ 
   \int \ d\vec{r} \
          e^{i \vec q \cdot \vec r (A-1)/A} \ 
          \rho^{(p)}(\vec r)
   \nonumber \\ &&
   \ + \ 
   f_{n}(q^2) \ 
   \int \ d\vec{r} \
          e^{i \vec q \cdot \vec r (A-1)/A} \ 
          \rho^{(n)}(\vec r)
   \>.
\end{eqnarray}
Here, $\rho^{(p)}(\vec r)$ and $\rho^{(n)}(\vec r)$ 
are the proton and neutron ground state one-body densities, 
which include corrections due to $2p2h$, $3p3h$, and $4p4h$ correlations. 

Similarly, we can write the second term as double integral over
the ground state two-body density, using
\begin{equation}
   \rho(\vec r_1,\vec r_2)
   \ = \
   \sum_{m n} \ 
   \langle
       \tilde 0 \, | \, \delta(\vec r_1 - \vec r_m) \ 
                        \delta(\vec r_2 - \vec r_n) \ 
       | \, \tilde 0
   \rangle
   \> .
\end{equation}
Then, the second term of Eq.~(\ref{eq:ffmbexp}) becomes
\begin{eqnarray}
   A_2 
   & = &
   f_{p}(q^2) \ 
   \int \ d\vec{r} \
   \int \ d\vec{r'} \
          e^{i \vec q \cdot \vec r (A-1)/A} \ 
          f^*(\vec q \cdot \vec r' / A) \ 
   \nonumber \\ && \hspace{0.75in}
          \left [ \rho^{(p,p)}(\vec r , \vec r')  
                  \, + \, 
                  \rho^{(p,n)}(\vec r , \vec r')  
          \right ]
   \nonumber \\ &&
   \ + \ 
   f_{n}(q^2) \ 
   \int \ d\vec{r} \
   \int \ d\vec{r'} \
          e^{i \vec q \cdot \vec r (A-1)/A} \ 
          f^*(\vec q \cdot \vec r' / A) \ 
   \nonumber \\ && \hspace{0.75in}
          \left [ \rho^{(n,p)}(\vec r , \vec r')  
                  \, + \, 
                  \rho^{(n,n)}(\vec r , \vec r')  
          \right ]
   \>.
\end{eqnarray}
With these evaluations we include all the terms that were included
in evaluating the one- and two-body densities.

%
%
\section{Results and Conclusions}

The problem of center of mass corrections in calculating observables
has been worked out by expanding the center-of-mass correction as
many-body operators. 
We have applied this expansion to the case of
the harmonic oscillator where an exact solution exists. We found
reasonable convergence in the case of harmonic oscillator wave
functions. 
Thus we have confidence that this method can be applied to
general Hartree-Fock wave functions and in a situation where 
2\emph{p}2\emph{h}-correlations are present.

\begin{figure}
   \centering
   \epsfxsize = 3.0in
   \subfigure[Argonne $v$18]
   {\epsfig{file=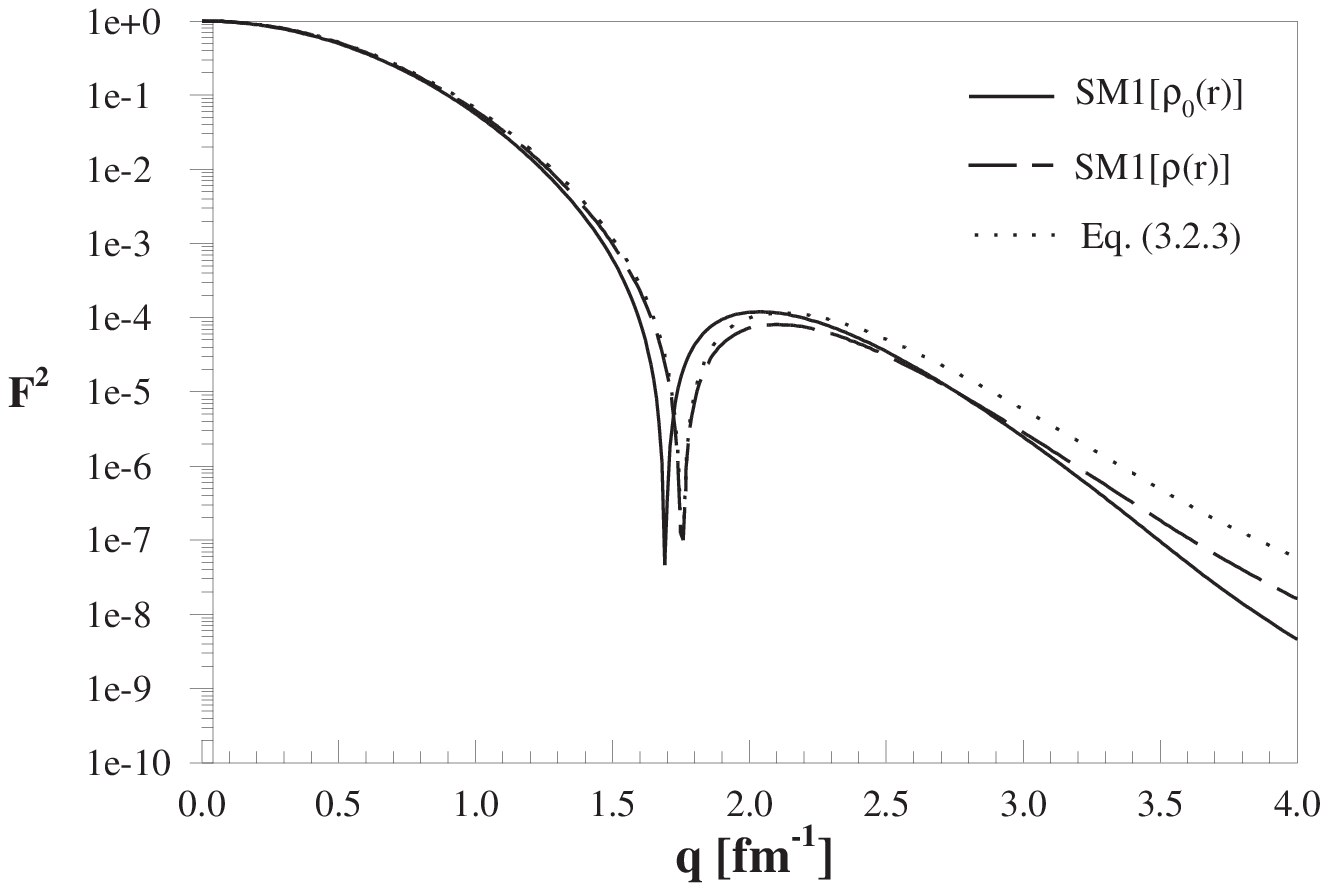,width=.50\textwidth}}
   \subfigure[Argonne $v$18 and Urbana-IX (preliminary)]
   {\epsfig{file=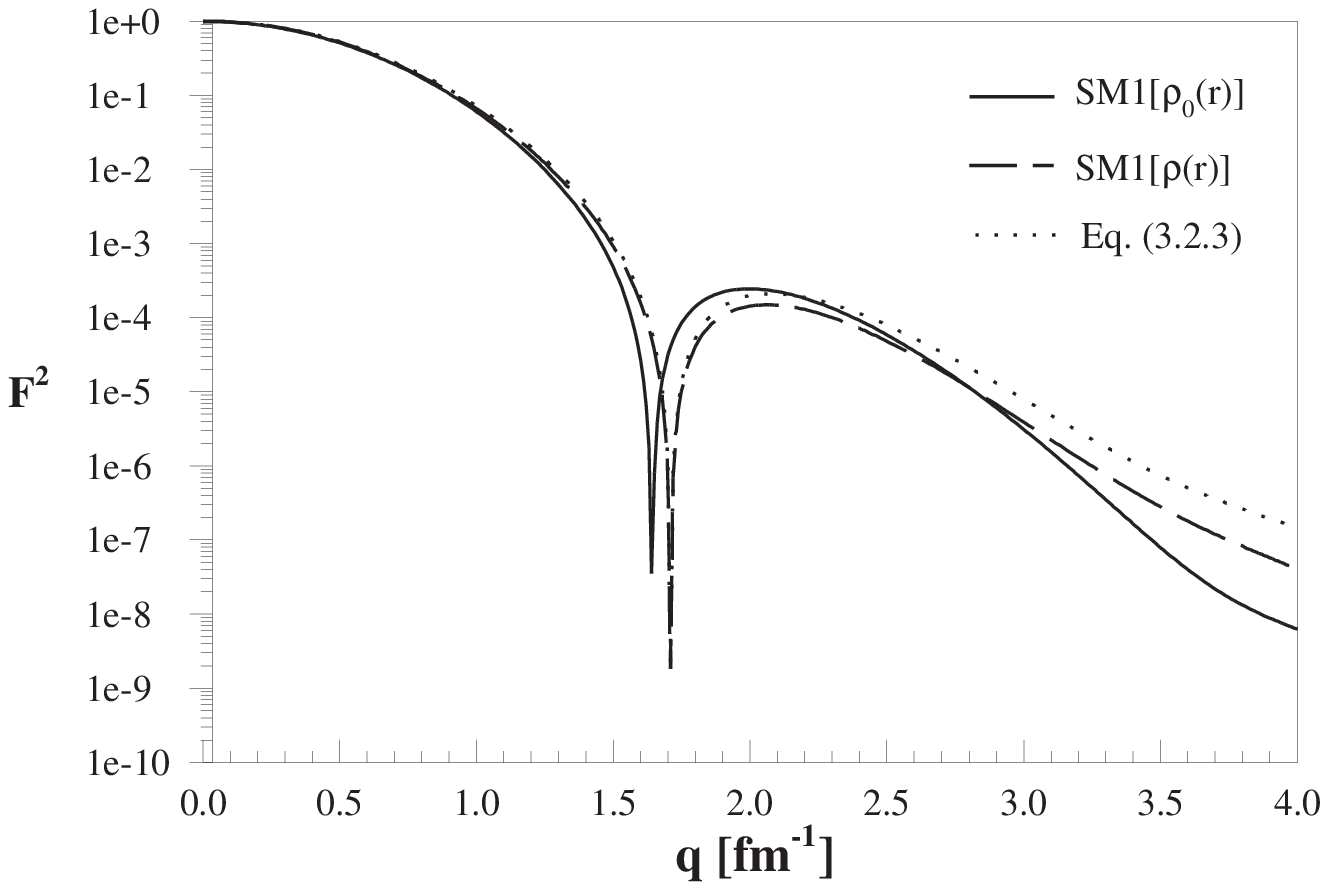,width=.50\textwidth}}
   \caption{$^{16}$O nucleus: 
            $SM1[\rho_0(r)]$ and $SM1[\rho(r)]$ form factors compared 
            with the internal form factor calculated according to 
            Eq.~(\ref{eq:ffCMcorr}).}
\label{fig:ff_nocor}
\end{figure}

\begin{figure}
   \centering
   \epsfxsize = 3.0in
   \subfigure[Argonne $v$18]
   {\epsfig{file=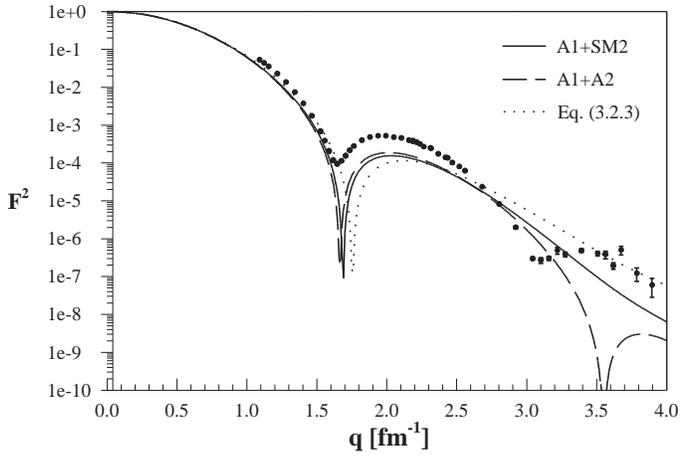,width=.50\textwidth}}
   \subfigure[Argonne $v$18 and Urbana IX (preliminary)]
   {\epsfig{file=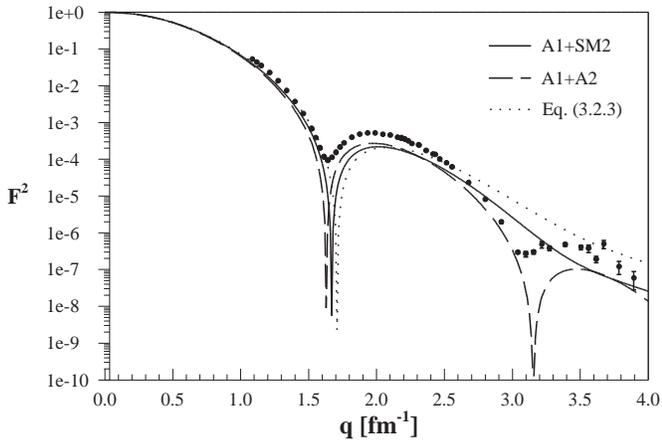,width=.50\textwidth}} 
   \caption{$^{16}$O nucleus: 
            Two-body approximations of the translational invariant form factor 
            compared with the internal form factor calculated according to 
            Eq.~(\ref{eq:ffCMcorr}).}
   \label{fig:ff_cor}
\end{figure}

Figures~\ref{fig:ff_nocor} and~\ref{fig:ff_cor} show 
the various effects of the correlations on the internal charge form factor,
corresponding to calculations using the Argonne $v$18
with/without the Urbana-IX potential .
We also compare the various approximations of the form factor
with the internal form factor suggested by Eq.~(\ref{eq:ffCMcorr}), 
which in both cases is plotted as a dotted line.

In the calculation of the translational invariant 
charge form factor correlations enter at two places. 
First, the calculation of the one-body operator (A1) includes effects 
of all the correlations, because this term is simply the Fourier transform 
of the one-body density.
In Fig.~\ref{fig:ff_nocor}, the solid and dashed lines represent 
the Fourier transform of the one-body density corresponding 
to the uncorrelated ($| 0 \rangle$) and correlated 
($| \tilde{0} \rangle$) ground state, respectively. 
These form factors are denoted $SM1[\rho_0(r)]$ and $SM1[\rho(r)]$.
Here, the main effect of the correlations is the shifting 
of the diffraction minimum by 5 \% to the right. The new minimum is also 
predicted by Eq.~(\ref{eq:ffCMcorr}), which also has a higher tail 
compared to $SM1[\rho_0(r)]$ and $SM1[\rho(r)]$.

Secondly, as any expectation value taken in the correlated ground state,
the center-of mass corrections are modified due to the correlations. 
In Fig.~\ref{fig:ff_cor}, the solid and dashed lines represent 
the two-body approximations of the translational invariant form factor.
Going beyond the leading order ($SM2$) in evaluating 
the two-body term ($A2$), leaves the first diffraction minimum 
virtually unchanged. 
However, the high $q$ behaviour of the form factor, ($q > 2.5$fm$^{-1}$), 
is dramatically affected.
We can see that the $A_1+A_2$ approximation of 
the internal charge form factor exhibits a second diffraction minimum,
which has been observed experimentally by Sick and McCarthy~\cite{ref:SickMcCarthy} and its presence makes our theory credible.
Physically speaking, the hole in the two-body density affects 
the center of mass motion and thus the center of mass correction 
to be applied.

%
%
\section*{acknowledgements}

This work was supported by the U.S. Department of Energy
(DE-FG02-87ER-40371).
Calculations were carried out on a HP-9000/735 workstation
at the Research Computing Center,
and a dual-processor 200 MHz Pentium Pro PC at the
Nuclear Physics Group of the University of New Hampshire.

%
%

\appendix

%
%
\section*{Notations.}
\label{sec:A}

We present here the various notations used in text. We have
\begin{eqnarray}
   &&
   I^{(1) \, L}_{\scriptstyle n_1 l_1 j_1, n_2 l_2 j_2}
   \ = \ (2 j_1 \,\delta_{(n_1 l_1 j_1), (n_2 l_2 j_2)} \, + \, 1)
   \nonumber \\ && \times
   \int_0^\infty 
   {\mathcal HO}_{n_1 l_1}(r) 
   {\mathcal HO}_{n_2 l_2}(r) 
   j_L(q r (A-1)/A) \, r^2 dr
   \>.
   \nonumber \\ 
\label{eq:I1}   
   \\
   &&
   I^{(2) \, L}_{\scriptstyle n_1 l_1 j_1, n_2 l_2 j_2}
   \ = \ (2 j_1 \,\delta_{(n_1 l_1 j_1), (n_2 l_2 j_2)} \,  + \, 1)
   \nonumber \\ && \times
   \int_0^\infty 
   {\mathcal HO}_{n_1 l_1}(r) 
   {\mathcal HO}_{n_2 l_2}(r) 
   f_L(q r / A) \, r^2 dr 
   \>.
\label{eq:I2}
\end{eqnarray}
Here, the symbol $\delta_{(n_1 l_1 j_1), (n_2 l_2 j_2)}$ 
is one when the set of indices ($n_1 l_1 j_1$) and ($n_2 l_2 j_2$) 
represent the same single-particle wave function, 
and zero otherwise.
We also introduce 
\begin{eqnarray}
   \lefteqn{
   \bar I^{(1,2) \, L}_{\scriptstyle n_1 l_1 j_1, n_2 l_2 j_2}
   }
   \nonumber \\ &&
   \ = \ 
   \langle (l_1 {\scriptstyle \frac{1}{2}}) j_1 \, || \, 
   \hat{C}^{(L)} \, || \, 
   (l_2 {\scriptstyle \frac{1}{2}}) j_2
   \rangle \ 
   I^{(1,2) \, L}_{\scriptstyle n_1 l_1 j_!, n_2 l_2 j_2}
\end{eqnarray}
The reduced matrix element of 
the unnormalized spherical harmonic operator of rank $k$ is 
\begin{eqnarray}
   \lefteqn{
   \left \langle
         ( l_a {\scriptstyle \frac{1}{2}} ) \, j_a
         \, \left \| \,
         C^{(k)}
         \, \right \| \,
         ( l_b {\scriptstyle \frac{1}{2}} ) \, j_b
   \right \rangle
   }
   \nonumber \\
   & = &
   (-1)^{j_a + k + {\scriptstyle \frac{3}{2}}} \,
   \sqrt{ \frac{(2j_a+1)(2j_b+1)}{(2k+1)} } \,
   \langle j_a \, {\scriptstyle \frac{1}{2}}
           \: j_b \, {\scriptstyle \frac{1}{2}}
           \ | \ k \, 0
           \rangle
   \nonumber \\
\label{eq:CKjajb}
\end{eqnarray}
for $| l_1-l_2 | \leq k \leq l_1+l_2$ and
$| j_1-j_2 | \leq k \leq j_1+j_2$ , and zero otherwise.

%
%

\hyphenation{Gart-en-haus}

\newpage

%
%
%


\newpage

%
%
%
\begin{table}
   \caption{Convergence of the mean square charge radius 
            for the case of the $^{4}$He and $^{16}$O nuclei.}
   \centering
   \begin{tabular}{lcc} 
      \\
      Order of approximation                   & $^{4}$He  & $^{16}$O \\ 
      \tableline
      \\
      $\mathrm{HO1}$                           & 1.285979  & 2.250000 \\ 
      $\mathrm{HO1}+\mathrm{HO2}$              & 1.484927  & 2.371708 \\
      $\mathrm{HO1}+\mathrm{HO2}+\mathrm{HO3}$ & 1.484922  & 2.349467 \\
      \\ 
      exact value                              & 1.484924  & 2.349468
      \\
   \end{tabular}
\label{tab:rms_ho}   
\end{table}


\begin{references}
   \bibitem{ref:paperone}
       J.H.~Heisenberg and B.~Mihaila,
       nucl-th/9802029
       (1998). 
   \bibitem{ref:argonnev18}
       R.B.~Wiringa and V.G.J.~Stoks,
       Phys. Rev. C \textbf{51}, 38 (1995).
   \bibitem{ref:tnipot}
       J.~Carlson, V.R.~Pandharipande, and R.B.~Wiringa,
       Nucl. Phys. A \textbf{401}, 59 (1983).
   \bibitem{ref:TassieBarker}
       L.J.~Tassie and F.C.~Barker,
       Phys. Rev. \textbf{111}, 940 (1959).
   \bibitem{ref:Lipkin}
       H.J~ Lipkin,
       Phys. Rev. \textbf{110}, 1395 (1958).
   \bibitem{ref:GartenhausSchwartz}
       S. Gartenhaus and C. Schwartz,
       Phys. Rev. \textbf{108}, 482 (1957).
   \bibitem{ref:Gibsonetal}
       B.F.~Gibson, A.~Goldberg, and M.S.~Weiss,
       Nucl. Phys. \emph{A} \textbf{111}, 321 (1968).
   \bibitem{ref:Feshbach}
       H.~Feshbach, A.~Gal, and J.~H\"ufner,
       Ann. Phys. \textbf{66}, 20 (1971).
   \bibitem{ref:Uberall}
       H.~Uberall,
       \emph{Electron Scattering from Complex Nuclei}
       (Academic Press, New York, 1971), Vol. 1, p. 183.
   \bibitem{ref:BarkerTassie}
       F.C.~Barker and L.J.~Tassie,
       Il Nuovo Cimento \textbf{XIX}, 1211 (1961).
   \bibitem{ref:ElliottSkyrme}
       J.P.~Elliott and T.H.R.~Skyrme,
       Proc. Phys. Soc. A \textbf{66}, 977 (1954).
   \bibitem{ref:SchiavillaETal}
       R.~~Schiavilla, V.~R.~Pandharipande, and D.~O.~Riska,
       Phys. Rev. C \textbf{41}, 309 (1990).
   \bibitem{ref:SickMcCarthy}
       I.~Sick and J.S.~McCarthy,
       Nucl. Phys. A \textbf{150}, 631 (1970).
\end{references}
\end{document}